\input harvmac
\input mssymb
\input labeldefs.tmp
\writedefs
\overfullrule=0pt

\input epsf
\newdimen\xfigunit
\global\xfigunit=4144sp
\def\figmag{1}
\newdimen\x \newdimen\y
\def\pic(#1,#2)(#3,#4)#5{%
\global\x=-#3\xfigunit%
\global\advance\x by-#1\xfigunit%
\global\y=-#4\xfigunit%
\def\epsfsize##1##2{\figmag##1}%
\epsfbox{#5}%
}
\def\put(#1,#2)#3{{\advance\x by#1\xfigunit\advance\y by#2\xfigunit%
\x=\figmag\x%
\y=\figmag\y%
\rlap{\kern\x%
\raise\y\hbox{#3}}}}
\def\fig#1#2#3{
\xdef#1{\the\figno}
\writedef{#1\leftbracket \the\figno}
\nobreak
\par\begingroup\parindent=0pt\leftskip=1cm\rightskip=1cm\parindent=0pt
\baselineskip=11pt
\midinsert
\centerline{#3}
\vskip 12pt
{\bf Fig.\ \the\figno:} #2\par
\endinsert\endgroup\par
\goodbreak
\global\advance\figno by1
}
\global\newcount\tabno \global\tabno=1
\def\tab#1#2#3{
\xdef#1{\the\tabno}
\writedef{#1\leftbracket \the\tabno}
\nobreak
\par\begingroup\parindent=0pt\leftskip=1cm\rightskip=1cm\parindent=0pt
\baselineskip=11pt
\midinsert
\centerline{#3}
\vskip 12pt
{\bf Tab.\ \the\tabno:} #2\par
\endinsert\endgroup\par
\goodbreak
\global\advance\tabno by1
}
\def\der{\partial}
\def\d{{\rm d}}
\def\e#1{{\rm e}^{#1}}
\def\E#1{{\rm e}^{\textstyle #1}}%

\def\Im{\mathop{\rm Im}\nolimits} 
\def\Z{{\Bbb Z}}
\def\C{{\Bbb C}}
\font\sans=cmss10
\def\D{{\sans D}}
\def\F#1{{\sans F}$_{#1}$}
\def\AF{{\sans AF}}
%
\def\pre#1{ (preprint {\tt #1})}
%
%
%
\lref\Lieb{E.~Lieb, {\it Phys. Rev. Lett.} 18 (1967), 692; 1046; 19 (1967), 108;
{\it  Phys. Rev.} 162 (1967), 162.}
\lref\Suth{ B. Sutherland, {\it Phys. Rev. Lett.} 19 (1967), 103.}
\lref\Kor{V.E. Korepin, {\it Commun. Math. Phys} 86 (1982), 391.}
\lref\Ize{A.G.~Izergin, {\it Sov. Phys. Dokl.} 32 (1987), 878.}
\lref\ICK{A.G.~Izergin, D.A.~Coker and V.E.~Korepin,
{\it J. Phys.} A 25 (1992), 4315.}
%
%
\lref\KZJ{V.~Korepin and P.~Zinn-Justin,
{\it J. Phys.} A 33 No. 40 (2000), 7053\pre{cond-mat/0004250}.}
\lref\PZJ{P.~Zinn-Justin,
{\it Phys. Rev.} E 62 (2000), 3411\pre{math-ph/0005008}.}
\lref\BBOY{M.T.~Batchelor, R.J.~Baxter, M.J.~O'Rourke and C.M.~Yung,
{\it J. Phys.} A28 (1995), 2759.}
\lref\JPS{W.~Jockush, J.~Propp and P.~Shor,
{\sl Random Domino Tilings and the Arctic Circle Theorem}
\pre{math.CO/9801068}.}
%
\lref\LW{E.~H.~Lieb and F.~Y.~Wu in {\sl Phase transitions and critical
phenomena vol.~1} (editors C.~Domb and M.~S.~Green), 331.}
\lref\CKP{H.~Cohn, R.~Kenyon and J.~Propp, {\it Journal of the AMS}
14 (2001), 297--346\pre{math.CO/0008220}.}
%
\lref\Ken{R.~Kenyon
in {\sl Directions in mathematical quasicrystals},
CRM Monogr. Ser., 13, Amer. Math. Soc. (2000), 307\pre{%
{\sl The planar dimer model with boundary: a survey},
http://topo.math.u-psud.fr/$\sim$kenyon/papers/papers.html}.}
%
\lref\Des{N.~Destainville, {\sl Entropie configurationnelle
des pavages al\'eatoires et des membranes dirig\'ees}, 
PhD dissertation (1997), and references therein.}
%
\lref\Kup{G.~Kuperberg, 
{\sl Symmetry classes of alternating-sign matrices under one roof}\pre{%
math.CO/0008184}.}
\lref\CEP{H.~Cohn, N.~Elkies and J.~Propp, {\it Duke Math. J.} 85 (1996)\pre{%
math.CO/0008243}.}
%
\Title{
\vbox{\baselineskip12pt\hbox{{\tt cond-mat/0205192}}}}
{{\vbox {
\vskip-10mm
\centerline{The Influence of Boundary Conditions}
\vskip2pt
\centerline{in the Six-Vertex Model}
}}}
\medskip
\centerline{P.~Zinn-Justin}\medskip
\centerline{\sl Laboratoire de Physique Th\'eorique et Mod\`eles Statistiques}
\centerline{\sl Universit\'e Paris-Sud, B\^atiment 100}
\centerline{\sl 91405 Orsay Cedex, France}
\vskip .2in

We discuss the influence of boundary conditions on the continuum limit
of the six-vertex model by deriving a variational principle for the associated
height function with arbitrary fixed boundary conditions. 
We discuss its consequences using the known phase diagram of the six-vertex model.
In some particular cases we compute explicitly the corresponding partial
differential equations by means of the Bethe Ansatz.
\Date{05/2002}

\newsec{Introduction}
The six vertex model, a well-known integrable model of
two-dimensional statistical mechanics, has been solved with various
types of boundary conditions: in the original
solutions, with periodic boundary conditions (PBC)
\refs{\Lieb,\Suth}, then more recently with
anti-periodic boundary conditions \BBOY\ and ``domain wall'' boundary
conditions (DWBC) \refs{\Kor,\Ize,\ICK,\KZJ,\PZJ}. Interestingly enough, 
in the latter case, even ``bulk'' quantities turned out to be different than with
PBC. One possible way to understand this is to notice that 
the six-vertex model possesses at each vertex a constraint: 
equality of the number of incoming and outgoing arrows.
This conservation of arrows renders the usual theorems of
statistical mechanics which ensure the existence of a thermodynamic limit
with bulk quantities that are independent of the boundary conditions
inapplicable (some energies being infinite); and it
effectively creates a non-locality
of the degrees of freedom (or of the moves from an
algorithmic point of view) which is expected to create sensitivity
on boundary conditions.
This raises the general problem of the effect of boundary conditions
on the thermodynamic limit of the six-vertex model. Important questions
such as the computation of the bulk free energy, as well as of some
local quantities (local polarization) will be addressed here. We shall
try to give a qualitative understanding of the physical phenomena involved,
as well as some explicit calculations whenever they are possible.

We shall consider in this paper fixed boundary conditions (FBC) from which
most boundary conditions can be derived. As we shall see, the main idea is
that these boundary conditions induce
{\it local}\/ polarizations.\foot{FBC fix in particular the total 
polarization, but the statement is stronger.} These
are responsible for spatial phase separation: considering
a continuum limit with a proper scaling for the domain in which
the model is defined and for the limiting conditions at its boundary,
we shall see that several phases may coexist in this domain.
Depending on the phase diagram (that is on the value
of the Boltzman weights defining the model),
the phases have definite boundaries in the case of second order
phase transition, whereas in the case of first order the phases freely mix.

The techniques we shall use are strongly inspired by the recent
developments in the field of dimers / domino tilings \refs{\JPS,\Ken,\CEP,\CKP}
which are themselves related to general ideas in random tilings \Des. As
was pointed out in this context in \refs{\KZJ,\PZJ}, 
domino tilings are a particular case
of the six-vertex model, corresponding to a special set of values of
the Boltzman weights.

The plan of the article is as follows. In section 2, we shall
define the six-vertex model with general FBC and
try to justify a conjectured variational principle for the
description of its thermodynamic limit. Then we shall analyze
the phase diagram
of the model in the three regimes that it possesses
and give explicit equations for the variational problem in some
cases (section 3).
Finally, we shall conclude in section 4 with some comments on the
applications of this variational principle (in particular
in relation to conjectures made in \refs{\KZJ,\PZJ}),
and some open questions.

\newsec{The six-vertex model and its variational principle}
The six-vertex model is defined on a regular square lattice. For fixed
boundary conditions (FBC) the lattice will be contained in a domain ${\cal D}$
of the plane
which is assumed, for simplicity, to be {\it convex}. For technical
reasons which will become clear below, we must also consider in parallel
the model with periodic boundary conditions (PBC), in which case
the lattice is on a torus. The lattice spacing is called $\delta$.

The configurations of the model are obtained
by assigning arrows to each edge of the lattice
(see Fig.~\config). The FBC mean that the arrows at the boundary
of ${\cal D}$ are supposed to be fixed once and for all. 
\fig\config{A configuration of the six-vertex model.}{\epsfxsize=4.5cm
\epsfbox{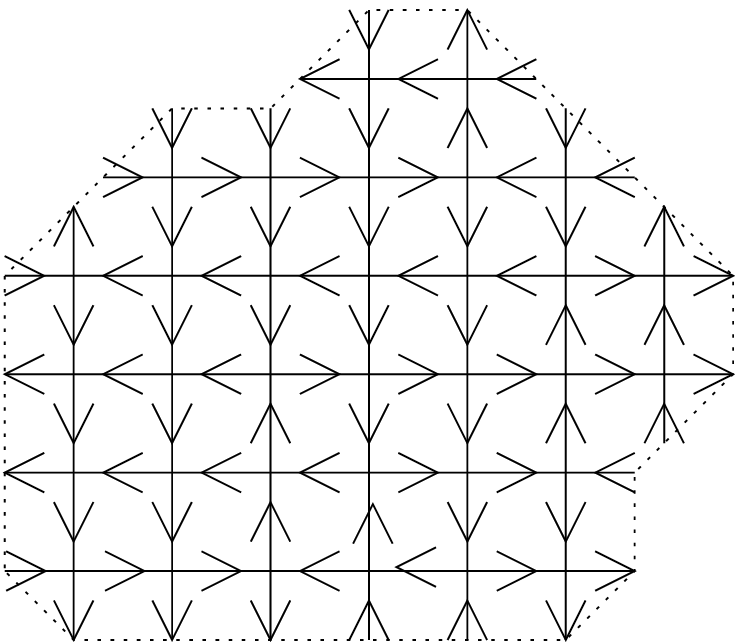}} 
The partition function
is then obtained by summing over all possible configurations:
\eqn\part{
Z=\sum_{\hbox{arrow configurations}}\ \prod_{{\rm vertex}\ v\in D} w(v)
}
where the statistical weights $w(v)$ are assigned to each {\it vertex} $v$
of the lattice. These are given in terms of the arrows around the vertex
according to
\eqn\wei{
w=\cases{
a\qquad\vcenter{\hbox{\epsfxsize=3.5cm\epsfbox{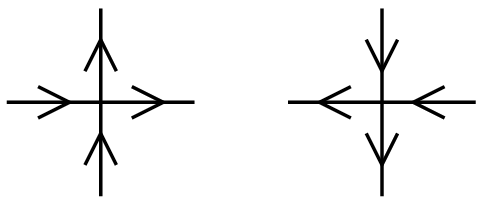}}}\cr
b\qquad\vcenter{\hbox{\epsfxsize=3.5cm\epsfbox{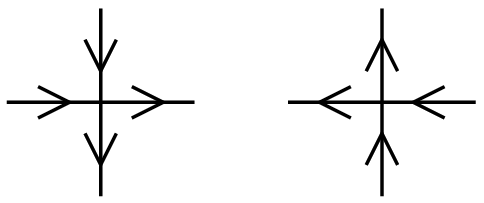}}}\cr
c\qquad\vcenter{\hbox{\epsfxsize=3.5cm\epsfbox{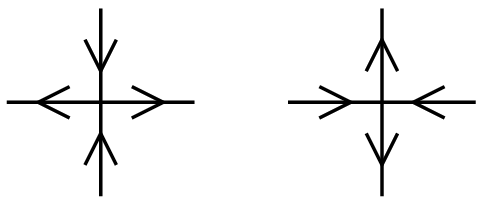}}}\cr
}
}
where $a$, $b$, $c$ are positive real numbers.
These being the only
configurations that respect the conservation of arrows, all the
other weights are zero.

One can consider the arrows as vectors in the plane with unit length;
then the {\it polarization}\/ of a subdomain ${\cal D}'\subset {\cal D}$ is defined as
$$\vec{P}({\cal D}')={1\over {\cal N}({\cal D}')}
\sum_{{\rm edge}\ e\in {\cal D}'} \vec{p}(e)$$
where ${\cal N}({\cal D}')$ is the number of vertices in ${\cal D}'$, and $\vec{p}(e)$ is
the arrow of edge $e$. Note that with FBC, the
total polarization per vertex $\vec{P}\equiv\vec{P}({\cal D})$ 
is fixed, but not with PBC. In the latter
case it is natural to introduce an electric field $\vec{E}$ coupled to
the polarization:
\eqn\partE{
Z(\vec{E})
=\sum_{\hbox{arrow configurations}}
\ \exp\Big(\vec{E}\cdot\sum_{{\rm edge}\ e\in {\cal D}}\vec{p}(e)\Big)
\ \prod_{{\rm vertex}\ v\in {\cal D}} w(v)
}
(we set for now the temperature to one. it will be restored when
needed). In the case of PBC it is known that the thermodynamic limit is 
well-defined; that is, independently of the shape of the torus, when one sends
its linear sizes to infinity there is a unique free energy per vertex
\eqn\defF{
F(\vec{E})=- \lim_{{\cal N}\to\infty}{1\over {\cal N}}\log Z(\vec{E})
}
where ${\cal N}={\cal N}({\cal D})$ is the number of vertices of the lattice.
One also defines for future use the Legendre transform of $F$:
\eqn\defG{
G(\vec{P})=F(\vec{E})+\vec{E}\cdot\vec{P}
}
where $\vec{P}$ and $\vec{E}$ are related by
\eqn\conj{
\langle \vec{P}\rangle=-{\der F\over \der \vec{E}}
}
As is well-known, $G$ is the free energy per vertex at fixed 
total polarization.

In the case of FBC, one also defines
the free energy per vertex to be
\eqn\GFBC{
G_{\rm FBC}=-{1\over {\cal N}}\log Z_{\rm FBC}
}
where the subscript FBC reminds us that $G$ depends on a choice
of the arrows at the boundary of ${\cal D}$.

Finally, we need to define the {\it height
function} $h$ associated to a configuration of the six-vertex
model with FBC; it is a function defined
on the faces of the lattice, such that when one moves from a face to
one of its neighbors, $h$ increases by $\delta$ if the arrow in between points right
(and of course decreases by $\delta$ if it points left). $h$ is only defined up to
a global constant. 
Similarly, there is a boundary height function
$h_0$ defined as above on the faces surrounding ${\cal D}$, and which is
fixed up to a constant
when the boundary arrows are fixed. In the end, the whole model can
be recast as a height model: the configurations are now $a\Z$-valued functions $h$
defined on the faces of the lattice, which satisfy $|h_f-h_{f'}|=\delta$
for neighboring faces $f$ and $f'$,
and the boundary condition $h_{|\der {\cal D}}=h_0$. The Boltzman weights are the direct
translation of those of the original model.

Now we are ready to take the continuum limit for FBC.
It is obtained by sending the lattice spacing
$\delta$ to zero, keeping the domain ${\cal D}$ fixed. What happens to
the boundary conditions? Since they can be specified by the boundary height
function $h_0$, we assume that the latter
converges to a given function $h_0$ defined on the boundary $\der {\cal D}$.
It is important to notice that typical configurations of the height function
will also be described by smooth functions defined on ${\cal D}$. In fact the constraint
$|h_f-h_{f'}|=\delta$ guarantees that any limiting height function will satisfy
the 1-Lipschitz condition $|h(x,y)-h(x',y')|\le|x-x'|+|y-y'|$, where $x$ and $y$ are
coordinates along the directions of the lattice, so that $h$ will be
(almost everywhere) differentiable.

More precisely, let us consider the average height
function
$\left<h(x,y)\right>$, which plays the role of one-point correlation
function in the model. In the continuum limit,
$\left<h\right>$ becomes a {\it macroscopic}\/ quantity.
In fact, its derivative is nothing but the average
local polarization:
\eqna\deri
$$\eqalignno{
{\der\over\der x} \left<h(x,y)\right>&=\left<P_y(x,y)\right>&\deri{a}\cr
{\der\over\der y} \left<h(x,y)\right>&=-\left<P_x(x,y)\right>&\deri{b}\cr
}$$
$\vec{P}(x,y)$ being defined by
$$\big<\vec{P}(x,y)\big>\equiv\lim_{{\cal D}'\ni(x,y)} \big<\vec{P}({\cal D}')\big>
$$
where ${\cal D}'$ is a macroscopic patch around $(x,y)$ whose size tends to zero.

The important consequence of this observation is that the summation over
$h$ in the partition function will be dominated by configurations near 
local minima due to a steepest descent phenomenon.
For the rest of this section we shall assume that we are in a non-degenerate
situation where the global minimum is unique and can be identified with
the average height function $\left< h\right>$ (we shall have to deal with
degenerate situations later on).

In order to go on we need an expression for the contribution to the free energy
of configurations near a given height function $h(x,y)$. We come to the main
hypothesis of this reasoning: we assume that the only effect of the FBC is to
create local polarizations. Thus, in every small patch ${\cal D}'$ 
of the domain ${\cal D}$ in which one can consider 
the local polarization $\vec P$ to be approximately constant, 
the free energy in ${\cal D}'$ is a function of $\vec P$ only. Now we already
encountered a model which is explicitly translationally invariant (i.e.\ 
with constant local polarization): the model with PBC, in which
one can constrain the polarization to the same value $\vec P$
by Legendre transform. 
We therefore conjecture the equality of free energies per vertex in these
two situations as we take the continuum limit.
There are many consistency checks of this hypothesis, some of which we shall see
below. The free energy being extensive, we reach the conclusion that
the contribution of configurations near $h(x,y)$ is given by
\eqn\contrib{
\int\!\!\!\int {\d x\d y\over \delta^2}\, G(-\der_y h(x,y),\der_x h(x,y))
}
where $G(P_x,P_y)\equiv G(\vec{P})$ is the free energy
per vertex with PBC at fixed polarization $\vec P$ defined earlier.
Note again that since $h$ is assumed to be $1$-Lipschitz
it is almost everywhere differentiable and the integral has a meaning. 

Eq.~\contrib\ is all we need to provide our variational principle. It is clear
now that the typical height function $\left<h\right>$ is simply obtained by
minimizing \contrib\ over all $1$-Lipschitz functions $h$ satisfying the
boundary condition $h_{|\der {\cal D}}=h_0$. 
And the free energy per vertex is given by
\eqn\varia{
G_{\rm FBC}=\min_{h} 
\int\!\!\!\int {\d x\d y\over {\cal A}({\cal D})} 
\, G(-\der_y h(x,y),\der_x h(x,y))
}
where ${\cal A}({\cal D})$ is the area of the domain ${\cal D}$. 
Eq.~\varia\ is our key formula.

Minimizing the functional above leads to interesting questions.
Indeed it is known that if $G(P_x,P_y)$ is a twice
differentiable {\it strictly} convex function
of $P_x$ and $P_y$ then $h(x,y)$ is the unique
solution of an elliptic partial differential equation (PDE), namely: 
\eqn\pde{
{\der^2 G\over\der P_y^2} (-\der_y h,\der_x h) \,  \der_{xx} h 
-2{\der^2 G\over\der P_x \der P_y}(-\der_y h,\der_x h)
\, \der_{xy} h 
+{\der^2 G\over \der P_x^2}(-\der_y h,\der_x h) \, \der_{yy} h 
=0
}
It turns out that
for no region of the parameters of the model is it fully the case;
however the problems are more or less severe depending on the regime. 
The general principles of statistical mechanics ensure that
the function $G$ is convex, but there is the possibility
of singularities (in our case, as we shall see,
the endpoints $|P_x|=|P_y|=1$ as well as the special point $P_x=P_y=0$);
and of linear parts (corresponding to the mixing
of two phases at a first order transition). We are thus led
to the study of the phase diagram of the model, which will be the
object of the next section.

Let us conclude here by describing how the results for
periodic and anti-periodic boundary conditions mentioned in the
introduction can be recovered in a self-consistent way in our
formalism. With PBC, one can still define a height function on a fundamental
domain, say $(x,y)\in [0,L]\times[0,L']$; however, the boundary conditions
now read
\eqn\PBCFBC{
h(x=L,y)=h(x=0,y)+L P_y\qquad h(x,y=L')=h(x,y=0)-L' P_x
}
where the total polarization per vertex $\vec P$ is arbitrary.
The free energy is now given by Eq.~\varia, but where $h$ is subject
to the boundary conditions \PBCFBC.
Linear functions clearly satisfy these,
and, for given $P_x$ and $P_y$, minimize \varia\ due to the convexity
of $G$. This means that the local polarization is constant and we immediately
recover the free energy $G(\vec P)$; in particular the unconstrained free
energy $F$, obtained by minimizing over $\vec P$, is simply $G(0)$. Modifiying
slightly the argument leads to an identical result for {\it anti-periodic} or 
{\it free} boundary conditions. 

\newsec{Bethe Ansatz solution and phase diagram}
The six-vertex model being integrable, one might think that a closed
expression exists for the free energy $F(\vec E)$ of the six-vertex
model with PBC in an electric field. In fact, even though Bethe Ansatz
equations exist for an arbitrary field $\vec E$, only for $\vec E=0$ can they
be solved in the thermodynamic limit. We briefly review here
the Bethe Ansatz equations.

Consider the six vertex model with PBC on a rectangle of size ${\cal N}=N\times M$.
One always starts
by using a transfer matrix formulation of the PBC partition function:
\eqn\TM{
Z=\tr T^M
}
where $T$ is the usual transfer matrix, 
acting on $(\C^2)^{\otimes N}$ that is on rows of vertical
arrows of the lattice. The goal is to diagonalize $T$; the largest
eigenvalue will then provide us with the free energy in the thermodynamic
limit.
Due to the conservation
of arrows (and the PBC), the number $n$ of up arrows is independent of
the row and we
can fix it. $n$ being related to the vertical polarization by
$P_y=1-2n/N$, the vertical eletric field plays no role and can be set to
zero; so that the
resulting free energy $\tilde{F}=-\lim_{N,M\to \infty}{1\over NM}\log Z$
is a mixed function of $P_y$ and $E_x$,
related by Legendre transform to both functions $F(E_x,E_y)$ and
$G(P_x,P_y)$ defined earlier.

There are many equivalent diagonalization procedures. In the coordinate
approach (see for example \LW),
one makes an Ansatz on the wave function of the $n$ up arrows. It
is characterized by $n$ distinct 
momenta $k_i$; these have to satisfy constraints (Bethe Ansatz Equations,
BAE)
obtained by having one spin up circle
around the strip of width $N$, interacting
with all the other spins up:
\eqn\bae{
\E{i k_j N}=\prod_{\scriptstyle \ell=1\atop l\ne j}^n B(k_j,k_\ell)
\qquad j=1,\ldots,n}
where
\eqn\smat{
B(p,q)=-
{1+\exp(4E_x+i(p+q))-2\Delta \exp(2E_x+ip)
\over
1+\exp(4E_x+i(p+q))-2\Delta \exp(2E_x+iq)}
}
and
\eqn\defD{
\Delta={a^2+b^2-c^2\over 2 a b}
}
The eigenvalue of the transfer matrix is then given by
\eqn\eig{
T=\big(a\e{E_x}\big)^N \prod_{j=1}^n L(k_j)+\big(b\e{-E_x}\big)^N \prod_{j=1}^n M(k_j)
}
where
\eqna\LM
$$\eqalignno{
L(k)&={a/b+\exp(ik+2E_x)-2\Delta\over \exp(ik+2E_x)a/b-1}&\LM{a}\cr
M(k)&={b/a+\exp(-ik-2E_x)-2\Delta\over \exp(-ik-2E_x)b/a-1}&\LM{b}\cr
}$$

In order to write down the final formula for the free energy, it is
convenient to introduce some more notations. We set
\eqn\nota{
a=\e{-\epsilon_a}\quad
b=\e{-\epsilon_b}\quad
c=\e{-\epsilon_c}
}
and split the contribution to the polarization of an arrow into its
two endpoints, so that the energies of the six vertex configurations become
$$\eqalign{
\epsilon_1=\epsilon_a - E_x - E_y
\quad&\vcenter{\hbox{\epsfxsize=3.5cm\epsfbox{weia.eps}}}\quad
\epsilon_2=\epsilon_a + E_x + E_y\cr
\epsilon_3=\epsilon_b - E_x + E_y
\quad&\vcenter{\hbox{\epsfxsize=3.5cm\epsfbox{weib.eps}}}\quad
\epsilon_4=\epsilon_b + E_x - E_y\cr
\epsilon_5=\epsilon_c
\quad&\vcenter{\hbox{\epsfxsize=3.5cm\epsfbox{weic.eps}}}\quad
\epsilon_6=\epsilon_c\cr
}$$

Then as $M\to\infty$ one can write
\eqn\finF{
\tilde{F}=\lim_{N\to\infty}\min \left[
\epsilon_1-{1\over N}\sum_{j=1}^n \log |L(k_j)|,
\epsilon_4-{1\over N}\sum_{j=1}^n \log |M(k_j)|
\right]
}
where $n={N\over 2}(1-P_y)$, 
and the $k_j$ are solutions of the Bethe Ansatz Eqs.~\bae\ 
chosen to minimize $\tilde{F}$.

There is an obvious symmetry $P_y\leftrightarrow -P_y$, and in
what follows we shall always restrict ourselves to $P_y\ge 0$, that
is $0\le n\le N/2$. This upper bound on $n$ appears naturally in the BAE.
Several scenarios can occur: the largest eigenvalue of $T$ can be attained
by the trivial state with all spins up ($n=0$); it can correspond
to the number $n$ reaching its maximum value $n=N/2$; or it can correspond
to a state with an intermediate number $0<n<N/2$.
After Legendre transformation,
similar phenomena take place for $P_x$. We
are therefore led to the following definitions:

\noindent $\diamond$ There may be regions of $\vec E$ for which
the polarization is constant and satisfies $|P_x|=|P_y|=1$. This
corresponds to a frozen phase with all the arrows aligned: we
call it the ferroelectric phase (\F{}). We can distinguish which
vertex configuration (from 1 to 4) is favored by the subscript:
\F{1}, \F{2}, \F{3}, \F{4}.

\noindent $\diamond$ There may be regions of $\vec E$ for which
the polarization is constant and is zero: we call it the
antiferroelectric phase (\AF).

\noindent $\diamond$ Finally, in regions of $\vec E$ where the
polarization is non-constant, there is no particular order and
we call this the disordered phase (\D).

The phase diagram depends very much on the value of the parameter
$\Delta$ (defined by \defD),
and we shall examine three cases separately. Note that in what
follows, we shall always assume $a>b$ due to the symmetry 
between $a$ and $b$. More detailed information on the phase diagram
can be found in \LW.

\subsec{The regime $\Delta>1$}
This is one of the two ``low temperature regimes''. 
As in all other regimes, it seems impossible to derive exact expressions for
the free energy in arbitrary field. Instead, it is natural
to look for a low temperature expansion. Let us start at zero temperature,
that is $\Delta=+\infty$.
In order to draw the zero temperature phase diagram, it is sufficient to compare the
four energies and pick the lowest. This leads to Fig.~\phaseFz.
\fig\phaseFz{Phase diagram as $\Delta\to +\infty$.}{%
\def\figmag{0.8}%
\pic(2744,2744)(879,-2333){F0.pstex}%
\put(2341,209){$E_y$}%
\put(3691,-1141){$E_x$}%
\put(1200,-100){\F{4}}%
\put(1451,-1681){\F{2}}%
\put(3061,-1906){\F{3}}%
\put(2891,-200){\F{1}}%
\put(2300,-520){$\scriptscriptstyle {\epsilon_b-\epsilon_a\over2}$}%
\put(2480,-860){$\scriptscriptstyle {\epsilon_b-\epsilon_a\over2}$}%
\put(1550,-1100){$\scriptscriptstyle {\epsilon_a-\epsilon_b\over2}$}%
\put(1750,-1420){$\scriptscriptstyle {\epsilon_a-\epsilon_b\over2}$}%
}

There is no bulk entropy and the free energy per vertex is simply
$F=\min(\epsilon_1,\epsilon_2,\epsilon_3,\epsilon_4)$. The Legendre
transformation is highly degenerate since the whole regions \F{i},
$i=1,\ldots,4$ correspond to the points $|P_x|=|P_y|=1$.
The square $(P_x,P_y)\in [-1,1]\times[-1,1]$ is now divided into
two regions, $P_x-P_y \lessgtr 0$,\foot{Had we chosen $b>a$ it
would have been  $P_x+P_y \lessgtr 0$.} which correspond to the
two triple points of Fig.~\phaseFz\ and where the phases \F{1},
\F{2}, and \F{3} or \F{4} mix. In these regions
$G$ is linear, cf Fig.~\GFz.
\fig\GFz{Free energy as a function of the polarization at $\Delta=+\infty$.}{%
\pic(3240,1710)(811,-2851){GF0.pstex}%
\put(1711,-2851){\F{2}}%
\put(811,-1681){\F{4}}%
\put(1711,-1141){\F{1}}%
\put(2611,-1681){\F{3}}%
\put(3601,-1591){$G$}%
\put(4051,-1861){$P_x$}%
\put(3151,-1861){$P_y$}%
}
The expression of the free energy:
\eqn\GFexpr{
G(P_x,P_y)=\epsilon_a+{\epsilon_b-\epsilon_a\over2}|P_x-P_y|
}
leads, for FBC, to the minimization of the following quantity:
$$I=\int\!\!\!\int {\d x\d y}\, |\der_y h+\der_x h|$$
where we recall that $h$ is defined on a convex domain ${\cal D}$,
is fixed to be $h_0$ on the boundary of ${\cal D}$, and satisfies
$|\der_x h|\le 1$, $|\der_y h|\le 1$.

We make the change of variables: $u=x-y$, $v=x+y$
so that $I=\int\!\!\!\int {\d u\d v}\, |\der_u h|$.
The integral over $u$ is clearly minimum when $\der_u h$ has fixed sign.
Let us call $u_1(v)<u_2(v)$ the intersections of $\der {\cal D}$
 and the $v={\rm cst}$
line; we see
that for fixed boundary values $h_0(u_1(v))$ and $h_0(u_2(v))$,
the minimum is
$$I=\int \d v |h_0(u_2(v))-h_0(u_1(v))|$$
How many functions realize this minimum? One can always choose
$h$ to be linear in $u$ at fixed $v$:
$$h(u,v)={u-u_1(v)\over u_2(v)-u_1(v)} h_0(u_2(v))
-{u-u_2(v)\over u_1(v)-u_2(v)} h_0(u_1(v))$$
This function is 1-Lipschitz if $h_0$ is. However,
for a generic function $h_0$, this will not be the only choice,
and in fact there will be an infinity of possible
$h$; this is related to the non-strict convexity of $G$, which is
itself related to the fact that the phase transitions of Fig.~\phaseFz\ are
first order.

The final expression for the FBC free energy for FBC at $\Delta=+\infty$ is:
\eqn\finalGF{
G_{\rm FBC}=\epsilon_a+{\epsilon_b-\epsilon_a\over2}
\int{\d v\over {\cal A}({\cal D})} |h_0(u_2(v))-h_0(u_1(v))|
}
It is however important to understand that the degeneracy in the choice of $h$
is an artefact of zero temperature, and is
lifted by any finite value of $\Delta$. Indeed, let us now look at a typical phase
diagram when $1<\Delta<+\infty$ (Fig.~\phaseF).
\fig\phaseF{Phase diagram in the $\Delta>1$ regime.}{%
\def\figmag{0.8}%
\pic(2744,2744)(879,-2333){F.pstex}%
\put(2341,209){$E_y$}%
\put(3691,-1141){$E_x$}%
\put(1150,-50){\F{4}}%
\put(1451,-1681){\F{2}}%
\put(3111,-1956){\F{3}}%
\put(2791,-376){\F{1}}%
\put(1580,-450){\D}%
\put(2750,-1600){\D}%
}

There is a first order transition left between phases \F{1} and \F{2}
on a finite interval; however the triple points have been replaced with disordered
regions with second order phase transitions at the boundary. The result, at
the level of the Legendre transform, is that the two flat faces of $G(P_x,P_y)$
acquire some curvature which make them strictly convex in the two regions
$P_x-P_y \lessgtr 0$. There
remains only a one-dimensional flat valley between the two.
Thus, we are led to the conclusion that in each of the spatial regions 
which correspond to $P_x-P_y \lessgtr 0$ (which are determined, as
stated earlier, by the sign of $h_0(u_2(v))-h_0(u_1(v))$),
$h$ is either ``saturated'' ($|\der_x h|=1$ or $|\der_y h|=1$), or
the solution of an elliptic PDE. The zero temperature limit of these
equations is given by the first correction
to the free energy. Here we shall not attempt the calculation of
this correction
since it is fairly involved, and a similar (though technically easier) 
calculation
will be performed in the more interesting regime $\Delta<-1$.
Note finally, that in the (non-generic) case of boundary conditions
such that there are regions for which $h_0(u_2(v))=h_0(u_1(v))$,
there is of course a unique way of minimizing the free energy for
all $\Delta>1$, which corresponds to remaining in the flat valley
$(\der_x+\der_y)h=0$ (mixture of \F{1} and \F{2}).

\subsec{The regime $\Delta<-1$}
This is traditionally called the antiferroelectric regime,
but in an electric field a disordered phase and even ferroelectric phases
can occur.

We start once again with
the zero temperature limit, which corresponds to $\Delta\to-\infty$.
In this limit, and in zero field,
the summation over six vertex configurations is dominated by
the contribution of the ground state, which is, up to a reversal of arrows,
given by Fig.~\AFgs, that is purely made of type $c$ configurations.

\fig\AFgs{Antiferroelectric ground state.}{\epsfxsize=3.5cm%
\epsfbox{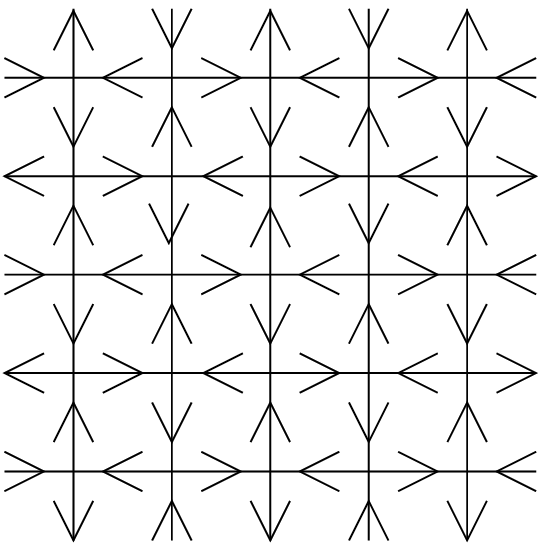}}%

The full phase diagram at $T=0$ is simply obtained, 
just as in the case $\Delta\to+\infty$,
by minimizing the energy of the six types of vertices, see Fig.~\phaseAFz.
\fig\phaseAFz{Phase diagram as $\Delta\to -\infty$.}{%
\def\figmag{0.8}%
\pic(2812,2744)(879,-2333){AF0.pstex}%
\put(2401,209){$E_y$}%
\put(3691,-1141){$E_x$}%
\put(1200,-100){\F{4}}%
\put(2431,-1321){\AF}%
\put(1451,-1681){\F{2}}%
\put(3061,-1906){\F{3}}%
\put(2850,-500){\F{1}}%
\put(1900,-1100){$\scriptscriptstyle {\epsilon_a-\epsilon_b\over2}$}%
\put(2360,-150){$\scriptscriptstyle{\epsilon_a+\epsilon_b\over2}-\epsilon_c$}%
}

We note that there are first order phase transitions from antiferroelectric
to ferroelectric phases. As before, we expect them to 
create degeneracies in the variational problem. In terms of the polarization
$\vec P$ (see Fig.~\GAFz), we have:
\eqn\GAFzeq{
G(P_x,P_y)=\epsilon_c+{\epsilon_a-\epsilon_c\over2}|P_x+P_y|+
{\epsilon_b-\epsilon_c\over2}|P_x-P_y|
}
\fig\GAFz{Free energy as a function of the polarization at $\Delta=-\infty$.}{%
\pic(2527,1834)(1059,-3493){GAF0.pstex}%
\put(3606,-3031){$P_x$}%
\put(3626,-2746){$P_y$}%
\put(3266,-2394){$G$}%
\put(2600,-2700){\F{1}}%
\put(1700,-1800){\F{4}}%
\put(750,-2550){\F{2}}%
\put(1620,-2700){\F{3}}%
\put(1700,-3180){\AF}%
}

The domain $\cal D$ will now be divided into {\it four}\/ subdomains 
depending on the sign of $(\der_x -\der_y) h$ and 
$(\der_x +\der_y) h$. In each of these subdomains, the free energy
being linear, there is generically an infinite number of functions
$h$ minimizing it and satisfying the constraints $|\der_x h|\le 1$,
$|\der_y h|\le 1$. The explicit determination of the
subdomains is however not as simple as
in the ferroelectric phase and can only be achieved on a case-per-case basis.

Let us now switch on the temperature.
When $-\infty<\Delta<-1$, as we can see on Fig.~\phaseAF, there is
an intermediate disordered phase which smooths the transitions
to second order.
\fig\phaseAF{Phase diagram of the $\Delta<-1$ regime.}{%
\def\figmag{0.8}%
\pic(2724,2724)(879,-2333){AF.pstex}%
\put(2341,209){$E_y$}%
\put(3621,-1141){$E_x$}%
\put(1200,-100){\F{4}}%
\put(2431,-1321){\AF}%
\put(1451,-1681){\F{2}}%
\put(3061,-1906){\F{3}}%
\put(2791,-376){\F{1}}%
\put(1060,-900){\D}%
}

It is convenient at this point to reintroduce a temperature $T=1/\beta$, so that the
weights become
\eqn\weiAF{
a=\e{-\epsilon_a/T}\quad
b=\e{-\epsilon_b/T}\quad
c=\e{-\epsilon_c/T}
}
Similarly to the regime $\Delta>1$, we expect the first low temperature 
correction
to lift the degeneracy and produce the PDE satisfied by $h$ as $T\to 0$
in the disordered regions.
Let us now turn to this calculation.

Because of the symmetries of the phase diagram \phaseAFz, we can choose to
expand around any of the four triple points. Note that these points are
not equivalent from the point of view of the Bethe Ansatz. We choose
$E_x$ near ${\epsilon_a-\epsilon_b\over 2}$; as a consistency check, 
if $P_y$ is
allowed to vary one should recover a value of $E_y$
near ${\epsilon_a+\epsilon_b\over2}-\epsilon_c$. 
Also, this corresponds to the region $|P_x|\le P_y$ 
(cf Fig.~\GAFz).

First we consider the simplification of the Bethe Ansatz in the limit $T\to 0$.
Assuming the $\exp(i k_j)$ to remain bounded in this limit, we immediately obtain
from Eq.~\smat\ that
$$B(p,q)\sim - \E{i(p-q)}$$
so that the Bethe Ansatz Eqs.~\bae\ have a straightforward solution.
Explicitly,
taking the logarithm and introducing half-integers $I_j$ (so that the $2I_j$ have
the opposite parity of $n$), we find
$$k_j(N-n)=2\pi I_j-K$$
where $K=\sum_j k_j$ is the total momentum, which is itself quantized:
$K=2\pi I/N$ with $I=\sum_j I_j$, so that
\eqn\solBAEAF{
k_j={2\pi\over N-n}(I_j-I/N)
}
The ground state is obtained by choosing the $k_j$ as close to zero as possible,
that is
\eqn\gsBAEAF{
k_j={2\pi\over N-n}\left(j-{n+1\over2}\right)\qquad j=1\ldots n
}
In particular the edge of the ``Fermi sea'' is $\pm Q$ where 
$$Q=\pi {1-P_y\over 1+P_y}$$

Next we consider the behavior of the energy functions \LM{} as $T\to 0$.
We introduce a scaling variable $u$ such that
$2E_x=\epsilon_a-\epsilon_b+ T u$; then Eqs.~\LM{} become
\eqna\LMAF
$$\eqalignno{
L(k)&={c^2/ab\over \e{u}\e{ik}-1}&\LMAF{a}\cr
M(k)&={c^2/ab\over \e{-u}\e{-ik}-1}&\LMAF{b}\cr
}$$
If $u>0$ $a\e{E_x/T}\gg b\e{-E_x/T}$ 
so that it is the first term in Eq.~\eig,
involving $L(k)$, that dominates; whereas if $u<0$ it is the second
term, involving $M(k)$.

Let us start with $u>0$. Using Eqs.~\finF, \gsBAEAF\ and \LMAF{} we find
\eqn\Fa{
\tilde{F}=\epsilon_1-T{n\over N}\int_{-Q}^{+Q} {\d k\over 2Q}
\log{c^2/ab\over \e{u}\e{ik}-1}
}
This is a dilogarithm integral; expanding in powers of $\e{-u}$ results in
\eqn\Fb{
\tilde{F}={1\over2}P_y(\epsilon_a+\epsilon_b-Tu)+(1-P_y)\epsilon_c
-T{1+P_y\over2\pi}\sum_{m=1}^\infty {\e{-mu}\over m^2} \sin(mQ)\quad u>0
}
Note that $E_y={\der \tilde{F}\over\der P_y}={\epsilon_a+\epsilon_b\over2}
-\epsilon_c+O(T)$, as expected.

If $u<0$, expanding $M$ in powers of $\e{u}$, one finds
the same expression but with $u$ replaced with $-u$.
It is a nice consistency check of our approach that these two expressions
are analytic continuations of each other as $u$ crosses zero. Indeed it is
clearly continuous at $u=0$, and the two explicit
expressions of its derivative
$$P_x=-{2\over T}{\der\tilde{F}\over\der u}=\cases{
P_y-{1+P_y\over\pi}\arg(1-\e{-u-iQ})\qquad u>0\cr
-P_y+{1+P_y\over\pi}\arg(1-\e{u-iQ})\qquad u<0\cr
}
$$
turn out to be equal.
In fact, the relation between $u$
and $\vec P$ can be rewritten much more conveniently as
\eqn\upol{
\e{u}={\sin\left(\pi{1-P_x\over 1+P_y}\right)
\over\sin\left(\pi{1+P_x\over 1+P_y}\right)}\qquad |P_x|\le P_y
}

When $u\to +\infty$, $P_x\sim P_y$, and taking the Legendre transform
of \Fa\ we obtain
\eqn\uinf{
G=P_y \epsilon_a + (1-P_y) \epsilon_c\qquad P_x=P_y
}
which is what is expected (first order phase transition between
phases \F{1} and \AF). A similar results holds for $u\to-\infty$.

Finally, noting that $u={\epsilon_b-\epsilon_a\over T}+2{\der G\over\der P_x}$,
we see that Eq.~\upol, supplemented by boundary condition \uinf, defines
entirely $G$. The explicit expression involves once again dilogarithms and
will not be written down here. What is more interesting is the second
derivatives of $G$, which provide the PDE \pde\ satisfied by $h$. After
a few calculations we find:
\eqna\finAF
$$\eqalignno{
{\der^2 G\over\der P_x^2}&={\pi\over2} {\sin{2\pi P_y\over 1+P_y}
\over (1+P_y)\sin{\pi(1 + P_x)\over 1 + P_y} \sin{\pi(1 - P_x)\over1 + P_y}}
&\finAF{a}\cr
{\der^2 G\over\der P_x\der P_y}&=-{\pi\over2}{\sin{2\pi P_x\over 1+P_y}-P_x\sin{2\pi\over 1+P_y}
\over (1+P_y)^2\sin{\pi(1 + P_x)\over 1 + P_y} \sin{\pi(1 - P_x)\over1 + P_y}}
&\finAF{b}\cr
{\der^2 G\over\der P_y^2}&=
{\pi\over2}{\left(\sin{2\pi P_x\over 1+P_y}-P_x\sin{2\pi\over 1+P_y}\right)^2
+\sin^2{\pi(1 + P_x)\over 1 + P_y} \sin^2{\pi(1 - P_x)\over1 + P_y}
\over (1+P_y)^3\sin{2\pi P_y\over 1+P_y}\sin{\pi(1 + P_x)\over 1 + P_y} \sin{\pi(1 - P_x)\over1 + P_y}}
&\finAF{c}\cr
}$$

One has ${\der^2 G\over\der P_x^2}{\der^2 G\over\der P_y^2}
-\left({\der^2 G\over\der P_x\der P_y}\right)^2={\pi^2\over (1+P_y)^4}$; 
the corresponding PDE, of the form \pde, is elliptic.

Note that Eq.~\finAF{} are only valid for $|P_x|\le P_y$;
similar expressions can be found in the other three regions by using
the various discrete symmetries of the model. These expressions
become singular when one approaches the boundaries $|P_x|=|P_y|$; this
is to be expected in a perturbative treatment around $T=0$, but such 
singularities
should disappear non-perturbatively. On the contrary, the singularity at
$P_x=P_y=0$ should remain due to the nature of the phase diagram (existence
of an antiferroelectric phase, cf Fig.~\phaseAF).

\subsec{The regime $-1<\Delta<1$}
This is traditionally called the disordered regime; however, as can be 
seen on the phase diagram (Fig.~\phaseD),
once the electric field is turned on both disordered and ferroelectric phases
occur.
\fig\phaseD{Phase diagram for $-1<\Delta<1$.}{%
\def\figmag{0.8}%
\pic(2724,2724)(879,-2333){D.pstex}%
\put(2341,209){$E_y$}%
\put(3691,-1141){$E_x$}%
\put(1200,-100){\F{4}}%
\put(1451,-1681){\F{2}}%
\put(3061,-1906){\F{3}}%
\put(2791,-376){\F{1}}%
\put(2000,-800){\D}%
}
The only difference with the phase diagram for $\Delta>1$ is that
there is no more phase boundary at $E_x=E_y=0$; instead the neighborhood
of this point is entirely inside the disordered phase.
The transitions from disordered to ferroelectric phases are second order.
This means that we expect that for given FBC, the domain ${\cal D}$ will be
divided into several zones (with definite boundaries),
each corresponding to a phase of Fig.~\phaseD:
typically the ferroelectric phases will lie close to the boundaries with
maximum slope ($|\der_x h_0|=1$ or $|\der_y h_0|$=1), where
their effect is strongest, while the disordered phase occupies the
remaining space. Inside the disordered zone, the function $h$
will satisfy an elliptic PDE.

Unfortunately, as in other phases this cannot be made explicit except in one
particular case: the ``free fermion point'', which will be described 
separately in the next section.
For a general value of $\Delta$ between $-1$ and $1$ the only exact result
that seems derivable 
is the low field expansion of the free energy. This is a standard computation
which will be sketched now.

First we introduce another parameterization of the weights:
\eqn\weipar{
a=\sin{\gamma\over2}(1-\zeta)\qquad
b=\sin{\gamma\over2}(1+\zeta)\qquad
c=\sin\gamma
}
with $-1<\zeta<1$ (or $-1<\zeta<0$ once we restrict ourselves to $a>b$), 
$0<\gamma<\pi$, so that $\Delta=-\cos\gamma$,
and make the standard change of variables in Eqs.\bae--\LM{}:
\eqn\chg{
\e{i k_j+2E_x}={\sinh \gamma(\lambda_j+i/2)\over\sinh\gamma(\lambda_j-i/2)}
}
The BAE become:
\eqn\baeD{
\left[{\sinh \gamma(\lambda_j+i/2)\over\sinh\gamma(\lambda_j-i/2)}
\e{-2E_x}\right]^N=
\prod_{\scriptstyle l=1\atop l\ne j}^n
{\sinh \gamma(\lambda_j-\lambda_\ell+i)\over\sinh\gamma(\lambda_j-\lambda_\ell-i)}
\qquad j=1,\ldots,n
}
and the functions $L$ and $M$ in terms of which the free energy is
express are given by
\eqna\LMD
$$\eqalignno{
L(\lambda)&={\sinh\gamma(\lambda-i(\zeta/2+1))\over\sinh\gamma(\lambda-i\zeta/2)}&\LMD{a}\cr
M(\lambda)&={\sinh\gamma(\lambda-i(\zeta/2-1))\over\sinh\gamma(\lambda-i\zeta/2)}&\LMD{b}\cr
}$$
Note that in the rhs of 
Eqs.~\baeD, only the differences $\lambda_j-\lambda_\ell$ appear.

Let us now assume $E_x=0$.
Then equations \baeD\ are real and their ``ground state'' solution is given
by real $\lambda_i$. If we further assume that
$n=N/2$ (that is after Legendre transformation,
$E_y=0$), then it is known that
the $\lambda_i$ fill the entire real axis in the limit $N\to\infty$, 
and their density is given by a simple integral equation. Namely, 
by taking the logarithm of \baeD\ and differentiating, we
find that the density $\rho_0(\lambda)$,
normalized by $\int\rho_0(\lambda)\d\lambda=n/N$, satisfies
\eqn\denseq{
K\star\rho_0 = \phi
}
where $\star$ denotes convolution product, and we give $K$ and $\phi$ directly
in Fourier transform: $\hat{K}(\kappa)=
\int K(\lambda) \exp(2i\kappa\lambda)\d\lambda
=2 {\sinh (\pi/\gamma-1)\kappa\cosh\kappa\over\sinh(\pi/\gamma)\kappa}$
and $\hat{\phi}(\kappa)={\sinh (\pi/\gamma-1)
\kappa\over\sinh(\pi/\gamma)\kappa}$,
so that we simply find $\hat{\rho}_0(\kappa)={1\over2\cosh\kappa}$ or
\eqn\denssol{
\rho_0(\lambda)={1\over2\cosh\pi\lambda}
}

If $n<N/2$, the equation remains similar, but
only an interval of the real axis, of the form $[-Q,Q]$,
is filled. The integral equation becomes equivalent to a $2\times 2$
Riemann--Hilbert problem which is in general unsolvable. However in the limit
$Q\to\infty$ (that is $P_y$, $E_y\to 0$), 
the effect of the interval endpoints on each
other become negligible are we are left with a simple Wiener--Hopf problem.
Let us briefly recall the principle. We concentrate on one of the two
edges, say $\lambda$ near $Q$, consider the new density $\rho(\lambda)$
and notice that its Fourier transform $\hat{\rho}(\kappa)\exp(-2i\kappa Q)$
is holomorphic in the half-plane $\Im \kappa<0$ and bounded by a polynomial
at infinity. The equation in Fourier transform is now:
\eqn\denseqb{
\hat{K}^{-1}(\kappa) \hat{\sigma}(\kappa)+\hat{\rho}(\kappa)=\hat{\rho}_0(\kappa)
}
where the unknown 
function $\sigma(\lambda)$ (the density of ``holes'') is, in our
approximation, such that
$\hat{\sigma}(\kappa)\exp(-2i\kappa Q)$ is on the contrary
holomorphic in the half-plane $\Im\kappa>0$. Solving this equation involves
the decomposition $\hat{K}(\kappa)=\hat{K}_+(\kappa)/\hat{K}_-(\kappa)$ where
\eqna\dechol
$$\eqalignno{
\hat{K}_+(\kappa)&={\Gamma(i\kappa/\gamma)\over
\Gamma((1/\gamma-1/\pi)i\kappa)\Gamma(1/2+i\kappa/\pi)}\e{\alpha\kappa}
&\dechol{a}\cr
\hat{K}_-(\kappa)&={i\over\pi}{\Gamma(1-(1/\gamma-1/\pi)i\kappa)\Gamma(1/2-i\kappa/\pi)
\over\Gamma(1-i\kappa/\gamma)}\e{\alpha\kappa}
&\dechol{b}\cr
}$$
and $\alpha=i(1/\gamma-1/\pi)\log(1-\gamma/\pi)+i/\pi\log(\gamma/\pi)$.
In general one must also decompose the right hand side of \denseqb,
but since $Q\to\infty$ one can consider only the contribution of the dominant
pole at $\kappa=-i\pi/2$, so that the solution of this analytic
problem is elementary:
\eqn\solrho{
\hat{\sigma}(\kappa)=\e{2iQ(\kappa+i\pi/2)}
{\hat{K}_+(-i\pi/2)\over \hat{K}_-(\kappa)} {i\over \kappa+i\pi/2}
}
$\hat{\rho}(\kappa)$ is undefined since in this $Q\to\infty$ limit,
$\hat{\rho}(\lambda+Q)$ diverges exponentially as $\lambda\to -\infty$; however
one can write
\eqn\solrhob{
\hat{\rho}(\kappa)=\e{2iQ(\kappa+i\pi/2)}
{\hat{K}_+(-i\pi/2)\over \hat{K}_+(\kappa)} 
{-i\over \kappa+i\pi/2}_{\big|\rm deformed}
}
where it is understood that to recover $\rho(\lambda)$ one must integrate
on a deformed contour that goes around the pole $\kappa=-i\pi/2$.

To recover the expression of $F$ as a function of $P_y$ we write, remembering
that there is also a contribution around $\lambda=-Q$:
\eqnn\polsol
$$
\eqalignno{
P_y&=2\int (\rho_0(\lambda)-\rho(\lambda))\d\lambda=-2\hat{\rho}(0)\cr
&={4\over\pi}\e{-\pi Q} {\hat{K}_+(-i\pi/2)\over \hat{K}_+(0)}&\polsol\cr
}$$
and
\eqnn\Fsol
$$
\eqalignno{
\tilde{F}&=\tilde{F}_0+2{\rm Re}
\int(\rho_0(\lambda)-\rho(\lambda))\log L(\lambda)\d\lambda\cr
&=\tilde{F}_0+2{1\over\pi}{\rm Res}_{\kappa=i\pi/2} \hat{\rho}(\kappa)\,
\widehat{\log L}(-i\pi/2)+\cdots
&\Fsol\cr
}
$$
so that after some calculations we find our first correction to the
free energy:
\eqn\Fsolb{
\tilde{F}_1={\pi\over 4}(1-\gamma/\pi)\cos(\zeta\pi/2) P_y^2
}

If we now switch on the horizontal field $E_x$, the factor $\exp(-2E_x)$
in Eqs.~\baeD\ breaks their reality and we can no longer assume that the
$\lambda_i$ are real; instead they will fill some curve in the complex plane.
At first order in $E_x$
the $\lambda_i$ will be displaced by a purely imaginary value,
so that we can write
$$\lambda_i=\lambda_i^0+iE_x u(\lambda_i)$$
where the $\lambda_i^0$ are the solutions of the BAE at $E_x=0$. Expanding the
BAE we easily find the equation for the function $\rho_u(\lambda)\equiv \rho(\lambda)
u(\lambda)$; in Fourier transform,
$$\hat{K}(\kappa) \hat{\rho}_u(\kappa)+\hat{\sigma}_u(\kappa)=2\delta(\kappa)$$
with the usual analyticity constraints on $\hat{\rho}_u$ 
and $\hat{\sigma}_u$. The solution is simply:
\eqna\solrhoc
$$\eqalignno{
\hat{\rho}_u(\kappa)&={-i\over \pi} \e{2iQ\kappa}
{1\over \kappa-i0} {\hat{K}_-(0)\over \hat{K}_+(\kappa)}&\solrhoc{a}\cr
\hat{\sigma}_u(\kappa)&={i\over \pi} \e{2iQ\kappa}
{1\over \kappa+i0} {\hat{K}_-(0)\over \hat{K}_-(\kappa)}&\solrhoc{b}\cr
}$$
This time the correction to the free energy is
\eqn\Fsolc{
\tilde{F}_2=-E_x +2iE_x\int\d\lambda \rho_u(\lambda) (\log L)'(\lambda)+O(E_x^2)
}
Rewriting the integral in Fourier transform, we see that its first two
dominant contributions come from the poles at $\kappa=0$ and $\kappa=i\pi/2$.
The pole at $\kappa=0$ compensates exactly the trivial term $-E_x$, and we are
left with:
\eqn\Fsold{
\tilde{F}_2= E_x P_y \sin(\zeta\pi/2)
}
Note that the absolute normalization of this term is meaningful.

The computation can be carried out to the next order in $P_y$; however
it is simpler to obtain it by $P_x\leftrightarrow P_y$ symmetry.
Combining Eqs.~\Fsolb\ and \Fsold\ and taking the Legendre transform, we
find:
\eqn\finD{
G(P_x,P_y)=G(0,0)+{\pi\over4}(1-\gamma/\pi) {1\over\cos(\zeta\pi/2)}
\left[ P_x^2+ 2 \sin(\zeta\pi/2) P_x P_y + P_y^2 \right]+\cdots
}
This is the low polarization expansion of $G$.

The corresponding linearized equation for $h$ is:
\eqn\lap{
\left({\der^2\over\der x^2}-2\sin(\zeta\pi/2){\der^2\over\der x\der y}
+{\der^2\over\der y^2}\right) h=0
}
This is a very simple result, which has the following interpretation:
for a low field we expect the usual conformal invariance to hold in the continuum
limit. It is known that the phase $|\Delta|<1$ is critical and its infrared limit
is described by a free massless boson. Here, $h$ plays the role of the
euclidean bosonic field.
Therefore it must satisfy $\Delta h=0$. However it is well-known 
that the spectral parameter $\zeta$ creates an anisotropy of the lattice: the
two directions of the lattice must be considered as forming an angle of
${\pi\over2}(1-\zeta)$, which leads to Eq.~\lap.

\subsec{The free fermion point $\Delta=0$}
When $\Delta=0$, that is when the Boltmann weights satisfy
the relation $a^2+b^2=c^2$, the system describes free fermions; this
is particularly clear when one considers the expression \smat\ of
the two-body interaction in the Bethe Ansatz equations, which becomes simply $B=-1$;
the BAE \bae\ themselves become
\eqn\freebae{
\E{i k_j N}=(-1)^{n-1}
}
i.e.\ describe (ignoring the sign issue related to our ``bosonic'' description
of fermions) the quantization of the momenta of free particles. The
$k_j$ still satisfy an exclusion principle, and the largest eigenvalue
of the transfer matrix is obtained by choosing
\eqn\freekj{
k_j={2\pi\over N}\left(j-{n+1\over2}\right)\qquad j=1\ldots n
}
The edge of the Fermi sea is
\eqn\freefs{
Q={\pi\over2}(1-P_y)
}
and the free energy is given by an integral of a type encountered above:
\eqn\freefe{
\tilde{F}=\epsilon_1-\int_{-Q}^Q {\d k\over 2\pi}
\log\left[{a/b+\e{ik+2E_x}\over \e{i k+2E_x}a/b-1}\right]
}

Finally, by taking the Legendre transform and differentiating twice,
we obtain the PDE (of the type \pde) 
satisfied by $h$ in the unsaturated regions
at $\Delta=0$:
\eqnn\PDEdz
$$\eqalignno{
0&=(a^2+b^2)\cos^2{\pi h_x\over2} \, h_{xx}\cr
&+2\left(a^2 \cos{\pi(h_x+h_y)\over2}-b^2\cos{\pi(h_x-h_y)\over2}\right)
\cos{\pi h_x\over2}\cos{\pi h_y\over2}\, h_{xy}\cr
&+(a^2+b^2)\cos^2{\pi h_y\over2}\, h_{yy}&\PDEdz\cr
}
$$
It is easy to check that this equation is elliptic for all $|h_x|<1$,
$|h_y|<1$.
If one sets $a=b$, 
after rotation of $\pi/4$ one recovers the PDE found in
\CKP\ for domino tilings.

\newsec{Conclusion}
We have proposed in this paper a variational principle (Eq.~\varia)
for the the six-vertex
model in the continuum limit. For generic fixed boundary conditions
this variational principle has a non-trivial solution, so that the existence
of a thermodynamic limit for the six-vertex model is invalidated.
A qualitative discussion of the phase diagram in the various regimes of the model
has suggested that in general the system undergoes phase separation
into regions which belong to one of three phases:
two ordered phases (ferroelectric, anti-ferroelectric), and one
disordered phase in which one can write a partial differential equation
satisfied by the height function (Eq.~\pde).
This leads to a first open question: it would be nice
to have a rigorous proof of the existence of such a phase separation.
In particular, it is not obvious at all in our framework
why an anti-ferroelectric region should exist at all, though we conjecture
it is generically the case for $\Delta<-1$. 
The calculations presented in this paper are essentially
a perturbative analysis around zero temperature, which is highly degenerate
and does not help to solve this issue. We have however given the explicit
PDE in the disordered regions in the limit $\Delta\to-\infty$ (Eqs.~\finAF).

It is unclear at the moment how useful this variational principle is in given
particular cases. One application is the domain wall boundary conditions
mentioned in the introduction. Note that for $\Delta=0$, $a=b$,
this is equivalent to the problem of domino tilings of the aztec diamond
\JPS, for which the height function is known exactly \CEP\ and solves
Eq.~\PDEdz. 
For $\Delta>1$, $a>b$, using the formalism of section 3.1, it is very easy to proof
the conjecture of section 4.1 (Fig.~4) of \KZJ. In our language, it
states that the
height function is given for all $\Delta>1$ by $h(x,y)=|x+y|$, $-1\le x,y\le1$
(i.e.\ it minimizes the free energy among the functions with identical
boundary values at $|x|=1$, $|y|=1$); it corresponds
to the non-generic case mentioned at the end of section 3.1 where one remains
in the flat one-dimensional valley $P_x=P_y$. Similarly, for $\Delta\to\infty$,
one can prove the conjecture of section 5.3 (Fig.~3) of \PZJ, with mild assumptions
on the possible form of the \AF/\F{} domains. One should however note that the DWBC,
with their maximum slope on all boundaries, are very non-generic boundary 
conditions, which explains in particular the absence of degeneracy for $\Delta=\pm
\infty$. Other similar boundary conditions which give rise to exact
determinant formulae are introduced in \Kup, and can probably be analyzed
along the lines of \KZJ\ or \PZJ; however, they presumably give rise to
the same height function, since they can be obtained from DWBC configurations
by dividing by a certain
discrete symmetry which is not spontaneously broken. It would therefore be useful
to find other exactly solvable boundary conditions to provide non-trivial
examples of the variational principle developed here.

\bigskip\centerline{\bf Acknowledgements}
The author would like to thank N.~Destainville and R.~Kenyon for useful
discussions.

\listrefs
\bye